\documentclass[journal=aamick,manuscript=article]{achemso}
\usepackage{chemformula} 

\title{PbTe/SnTe heterostructures -- candidate platform for studying  spin-triplet superconductivity
}

\author{P. Sidorczak}
\affiliation{University of Warsaw, Faculty of Physics, ul. Pasteura 5, 02-093, Warsaw, Poland.}
\alsoaffiliation{International Research Centre MagTop, Institute of Physics, Polish Academy of Sciences, al. Lotnikow 32/46, 02668,  Warsaw, Poland.}
\alsoaffiliation{Center for Quantum Devices, Niels Bohr Institute, University of Copenhagen,  Universitetsparken 5, 2100 Copenhagen, Denmark.}
\email{pawel.sidorczak@nbi.ku.dk} 
\author{W. Wo\l{}kanowicz}
\affiliation{Institute of Physics, Polish Academy of Sciences,  \small al. Lotnikow 32/46, 02668,  Warsaw, Poland.}
\author{A. Kaleta}
\author{M. W\'ojcik}
\author{S. Giera\l{}towska}
\author{K. Gas}
\affiliation{Institute of Physics, Polish Academy of Sciences,  \small al. Lotnikow 32/46, 02668,  Warsaw, Poland.}
\alsoaffiliation{Center for Science and Innovation in Spintronics, Tohoku University, Katahira 2-1-1, Aoba-ku, Sendai 980-8577, Japan.}
\author{T. P\l{}oci\'nski}
\affiliation{Faculty of Materials Science and Engineering, Warsaw University of Technology, ul. Woloska 141, 02-507, Warsaw, Poland.}
\author{R. Minikayev}
\affiliation{Institute of Physics, Polish Academy of Sciences,  \small al. Lotnikow 32/46, 02668,  Warsaw, Poland.}
\author{S. Kret}
\affiliation{Institute of Physics, Polish Academy of Sciences,  \small al. Lotnikow 32/46, 02668,  Warsaw, Poland.}
\author{M. Sawicki}
\affiliation{Institute of Physics, Polish Academy of Sciences,  \small al. Lotnikow 32/46, 02668,  Warsaw, Poland.}
\alsoaffiliation{Research Institute of Electrical Communication, Tohoku University Katahira 2-1-1, Aoba-ku, Sendai 980-8577, Japan.}
\author{T. Wojtowicz}
\affiliation{International Research Centre MagTop, Institute of Physics, Polish Academy of Sciences, al. Lotnikow 32/46, 02668,  Warsaw, Poland.}
\author{D. Wasik}
\affiliation{University of Warsaw, Faculty of Physics, ul. Pasteura 5, 02-093, Warsaw, Poland.}
\author{M. Gryglas-Borysiewicz}
\affiliation{University of Warsaw, Faculty of Physics, ul. Pasteura 5, 02-093, Warsaw, Poland.}
\email{Marta.Gryglas@fuw.edu.pl} 
\author{K. Dybko}
\email{dybko@ifpan.edu.pl}
\affiliation{Institute of Physics, Polish Academy of Sciences,  \small al. Lotnikow 32/46, 02668,  Warsaw, Poland.}
\alsoaffiliation{International Research Centre MagTop, Institute of Physics, Polish Academy of Sciences, al. Lotnikow 32/46, 02668,  Warsaw, Poland.}

\begin{document} 

\newpage\begin{abstract} 
This paper explores the potential for spin-triplet superconductivity in molecular beam epitaxy grown PbTe/SnTe semiconductor heterostructures. We present convincing evidence for spin-triplet pairing by soft point-contact spectroscopy experiments, using both spin-polarized and unpolarized electrons and additionally, by detailed analysis of the upper critical field, as inferred from the four probe resistance measurements. The experimental data are described in terms of the Anderson-Brinkman-Morel model of p-wave electron pairing. Our results confirm the predictions on strain-induced topological superconductivity by E.Tang and L. Fu (\em {Nature Physics}, 10, 964, 2014).
\end{abstract}

\subsection*{Introduction}
The search for spin-triplet superconductors is propelled by their potential applications in spintronics and quantum computation \cite{narlikar_superconducting_2017, linder_superconducting_2015, beenakker_road_2016, flensberg_engineered_2021, alicea_new_2012, sato_topological_2017,  sato_topological_2009, sato_topological_2010,Tanaka2024}. It is predicted that these superconductors (SCs) host Majorana fermions at their boundaries. Recent breakthroughs in twisted bi-layer graphene experiments have unveiled signatures of spin-triplet superconductivity at specific "magic angles" \cite{cao_pauli-limit_2021,oh_evidence_2021}, due to strain-induced electron flatbands. The exploration of a flatband regime presents exciting possibilities for engineering high-$T_c$ superconductors \cite{volovik_flat_2013, esquinazi_flat_2016, lau_designing_2021}.  Additionally, intriguing results have surfaced in heavy fermion compounds \cite{ran_nearly_2019,shimizu_spin-triplet_2019} and unconventional materials like K$_2$Cr$_{3}$As$_{3}$ \cite{yang_spin-triplet_2021}, challenging the conventional understanding of superconductivity. Sr$_2$RuO$_4$, once believed to be the sole confirmed case of a triplet superconductor, has faced recent reinterpretation \cite{pustogow_constraints_2019,abergel_triplet_2019}. Proximity-induced spin-triplet superconductivity, involving injecting spin-singlet Cooper pairs into a medium with ferromagnetism or spin-orbit coupling, has also been explored \cite{reeg_proximity-induced_2015,nakamura_evidence_2019,cai_superconductor/ferromagnet_2023}.
The p-wave symmetry of the Cooper  pairs has also been well identified and described in the superfluid helium-3.\cite{silaev_andreev-majorana_2014, Heikkil2016, Mizushima2016}.

Our study focuses on a different superconducting system: IV-VI
semiconductor heterostructures. First observations of the
superconductivity in those systems date back to the 80' \cite{murase_superconducting_1986}. Soon later presence of dislocations was evidenced
\cite{fedorenko_strain_1992} and their correlations to the superconductivity were
established \cite{fogel_novel_2001,fogel_interfacial_2002, fogel_direct_2006, yuzephovich_interfacial_2008,bengus_suppression_2013}. It was found that the
superconducting state develops at the interface of the strained layers.
The renaissance of topology in the condensed matter physics has brought new
insights into the foundations of this unusual superconductivity. E.Tang and
L. Fu proposed that a misfit dislocation array at the interface 
produces a periodically varying strain field and flatbands with a high
density of states at the Dirac points, giving rise to the interface
superconductivity observed in IV–VI compounds, which is also predicted  to exhibit
non-Bardeen–Cooper–Schrieffer (BCS) behaviour.\cite{tang_strain-induced_2014,hsieh_topological_2012} In parallel,
experimental evidence of strain in these systems has been presented \cite{Zeljkovic2015}. However, up to now, no one has experimentally investigated
the symmetry of electron pairing.

We used point-contact spectroscopy of PbTe/SnTe heterostructures to determine the pairing symmetry of the  strain-induced superconductivity. The preliminary results of spin polarized point-contact spectroscopy (spin-polarized PCS) pointed to spin-triplet symmetry. Systematic studies of differential conductance spectra were performed on a series of samples as a function of temperature and magnetic field. All the spectra revealed zero bias conductance peak (ZBCP), expected for the unconventional superconductivity. The spectra  are consistent with the Anderson-Brinkman-Morel (ABM) anisotropic p-wave superconductor model \cite{he_evidence_2022,soulen_andreev_1999,anderson_generalized_1960, leggett_theoretical_1975, anderson_anisotropic_1973}. This model was developed for the case of spin-triplet pairing identified in superfluid $^{3}$He \cite{feng_introduction_2005,coleman_introduction_2015}. Basic parameters of the superconducting state were extracted from resistance dependence on temperature and magnetic field, $R(T,H)$, pointing to its two-dimensional character, as predicted by Tang and Fu \cite{tang_strain-induced_2014,fogel_novel_2001,fogel_interfacial_2002,fogel_direct_2006}.

\subsection*{Results}

\subsubsection*{Growth and Structural information}

\begin{figure}[htbp]
\centering
\includegraphics[width=0.55\linewidth]{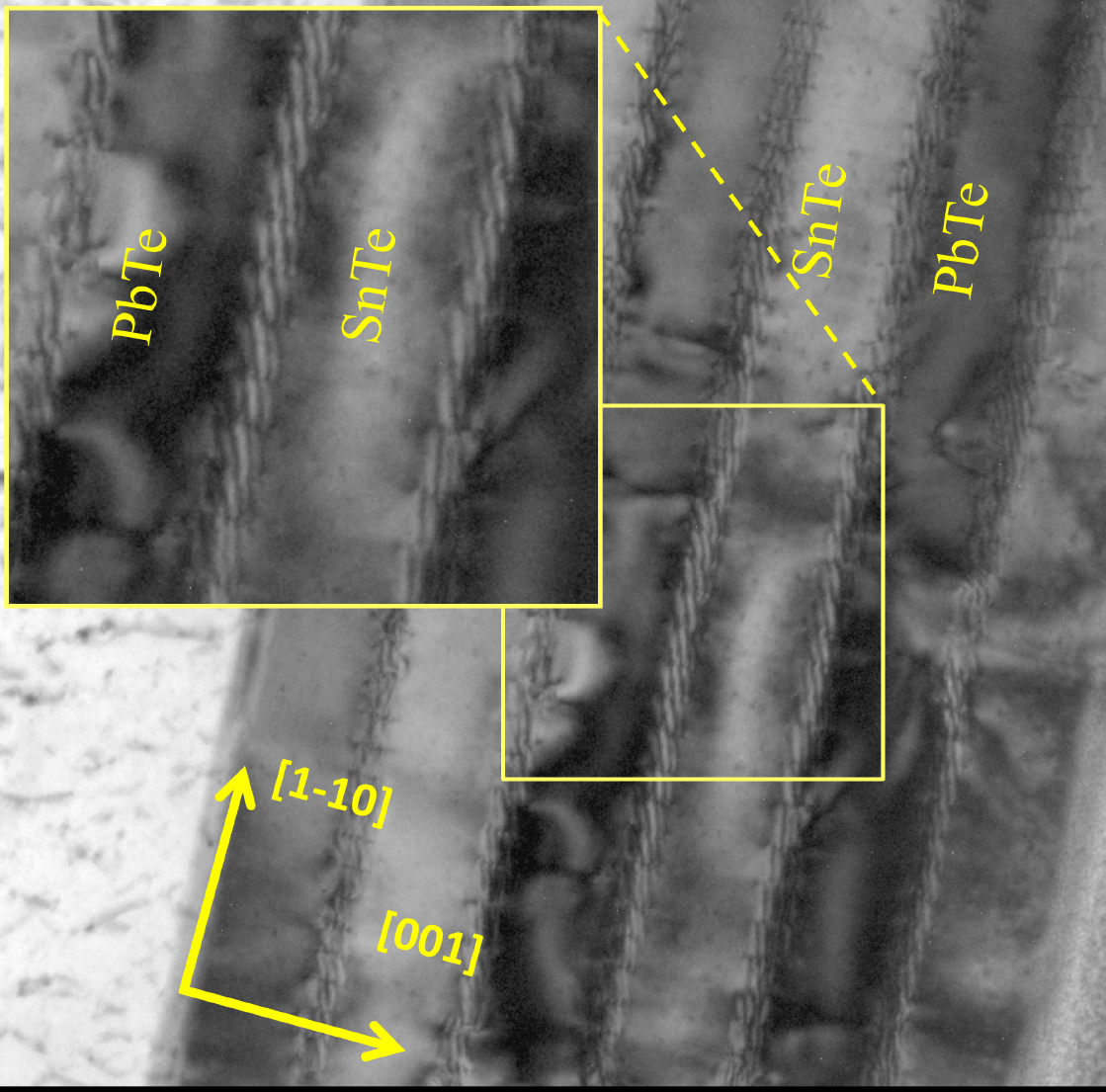}
\caption{\label{fig:intro}
	Results of electron transmission microscopy.
	 TEM image of the cross-section of PbTe/SnTe heterostructure. The alternating streaks along layer's interfaces mark periodic dislocation grid.}
\end{figure}

Our PbTe/SnTe semiconductor heterostructures are grown by molecular beam epitaxy (MBE) in temperature range of 270\textcelsius - 300\textcelsius \ using system with effusion cells containing Pb and Te as elements and a compound source of SnTe. Such combination allows us to precisely control
stoichiometry of deposited layers which were grown on KCl substrate and semi-insulating GaAs
substrate with 4$\mu$m thick CdTe buffer layer oriented both in [100] direction. The role of CdTe
buffer is to eliminate defects and strains appearing on the interfaces due to the great mismatch in
lattice parameters between GaAs (5.65 Å) and PbTe (6.46 Å) and provide plane surface for
PbTe/SnTe heterostructures. Additionally to ensure the buffer growth in [100] direction very thin
film of few monolayers of ZnTe was deposited as it is known to prevent the Stransky–Krastanov growth mode.
A two-dimensional mode of growth has been confirmed by streaky reflection high energy electron diffraction (RHEED) patterns. Lattice mismatch between two consecutive semiconductors leads to the formation of a periodic dislocation grid at the interface. It is visualized in Figure \ref{fig:intro} using the Bright Field transmission electron microscope (TEM) imaging condition obtained near [110] axis. Moreover, we underline that the Z-contrast scanning transmission electron microscope (STEM) shows the layers to be almost fully relaxed far from the interfaces and that no extended defects in the form of threading dislocations or stacking faults crossing the whole heterostructure are present. Heterostructures with up to 7 semiconductor layers are studied. Due to high technology of MBE growth we were able to exceed the $\text {T}_\text{c}$ of 6 K, while the earlier works reported the $\text{T}_\text{c}$ of 3K for this type of heterostructure. \cite{fogel_novel_2001} Collection of the extended RHEED, X-ray diffraction, and TEM characterization is presented in the Supporting Information  (Figures S1-S2).

\subsubsection*{Point Contact Spectroscopy}

\begin{figure}[htbp]
\centering
	\includegraphics[width=0.75\linewidth]{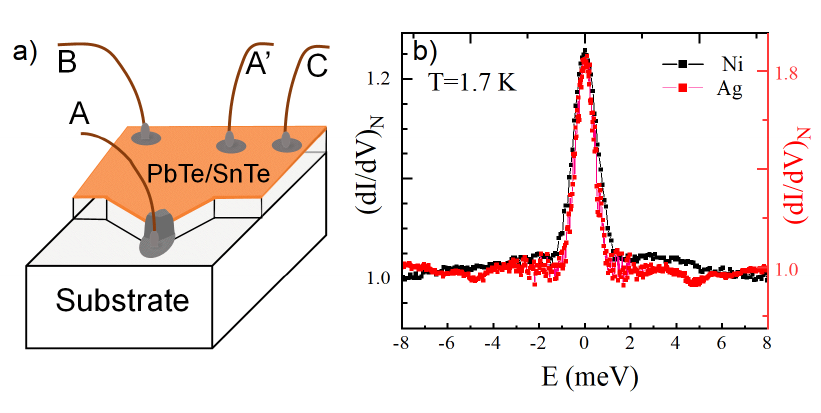}
\caption{\label{fig:intropcs}
Point contact spectroscopy.
	a) Scheme of contacts arrangement for differential conductance measurements: 
	i) voltage(B,A'), current(C,A') -- all contacts on top of PbTe layer,
	ii) voltage(B,A), current(C,A) -- tunneling contact A is attached to the cleaved side of heterostructure i.e., to the superconducting layer directly.
	b) Normalized \textit{dI/dV} spectra measured with both nickel (spin polarization of nickel $P_{Ni}\approx46\%$ \cite{nadgorny2000}) paste and silver ($P_{Ag}\approx0\%$) paste contacts at $T=1.8$ $K$. Spin polarized current preserves Andreev zero bias conductance peak, which is a fingerprint of spin-triplet superconductivity.}
\end{figure}

Point-contact spectroscopy is an experimental technique involving the Andreev reflection (or tunneling) of an electron on the interface between a superconductor and a normal metal. This is a process in which a single electron with momentum $k$ is injected into a superconductor and binds with a free electron forming a Cooper pair. Since the momentum, spin, and number of particles are conserved, a single hole is reflected with momentum equal to $-k$ and an adequate spin \cite{daghero_probing_2010}. 
This process can be studied by measuring differential conductance as a function of voltage bias $V$ corresponding to incident electron energy (eV), schematically shown in the Figure \ref{fig:intropcs}a. This technique also allows the determination  spin polarization of ferromagnetic contacts \cite{Soulen1998, Chen2010, Chen2012, Panguluri2004}, as well as to study the SC order parameter $\Delta$ and to distinguish between different orbital symmetries (s-wave, p-wave, d-wave)\cite{wang_tip-induced_2018}. It was  used for qualitative demonstration of topological superconductivity in $Cu_xBi_2Se_3$ \cite{Sasaki2011} as well as for quantitative analysis of p-wave spectra in PCS of $UTe_2$ \cite{Yoon2024-jz}.

In this work soft PCS is employed and silver paste contacts are used. \cite{daghero_probing_2010,wang_tip-induced_2018} 
 The results of PCS revealed that the spectra consist of a single zero-bias conductance peak (ZBCP) together with symmetric dips on the sides. Remarkably, such  results are independent of spin polarization $P$ 
 pointing to spin triplet SC, as shown (for different samples) in Figure \ref{fig:intropcs} and Figure S4 in Supporting Information. The two curves for spin-polarized and unpolarized current are similar, which stays in the stark contrast to the results of control experiment where Ni/Ag contacts are applied to s-wave superconductor, Niobium, (see Figure S6 in the Supporting Information), clearly showing that even partially spin polarized Ni paste may be successfully  used as a source of spin polarized carriers, leading to significant suppression of differential conductance at zero bias. 
 It is also worth to mention that spin polarization of injected carriers is preserved on its  whole path from the point contact on top layer PbTe to superconducting island, because the electron mean free path in PbTe is suffitiently long  ($l_{mfp} > 1 \; \mu \text{m} > d_{PbTe} = 100$ nm). 
 \cite{Prinz1999,Kolwas2012} 
Moreover, the strong spin--orbit coupling in PbTe and SnTe does not lead to spin depolarization because  the rock--salt symmetry of the crystal ensures the presence of an inversion symmetry center which in turn cancels out linear and cubic terms in wave vector responsible for spin--orbit scattering.

As quoted above, for the Andreev reflection to occur, there needs to be a hole available in the emitter with adequate spin in relation to the incident electron: opposite (singlet pairing) or identical (triplet pairing). If a current is injected to the SC via spin polarized contact, such as a ferromagnetic metal, the above condition will only be fulfilled for a spin-triplet superconductor \cite{he_evidence_2022}. In our case, the spectra are qualitatively identical, differing only in amplitudes. The latter can be ascribed to various grain sizes of  silver and nickel pastes, different broadening parameters, and variations in the Fermi energy between silver and nickel.  Moreover, as shown in Supporting Information Figure S5a) and b), we exclude the possibility that ZBCP is due to either the thermal regime of the point-contact or the Josephson junction coupling between superconducting domains \cite{daghero_probing_2010,he_distinction_2018}.

The results presented above form  a qualitative proof of unconventional superconductivity in PbTe/SnTe heterostructures. To gain more insight, the differential conductance spectra are analyzed with the model calculated for normal metal/ABM p-wave superconductor junction with a 4-component wave function \cite{he_evidence_2022,schlomer_nschloe/quadpy_2018,harris2020array}. Such a system can be characterized by the following parameters: $\Delta$ - order parameter, $Z$ - transparency of a $\delta$-like barrier between the SC and normal metal, $\Gamma$ - inelastic scattering factor, and $\phi$ - the angle describing the relative position of the x-axis of the p-wave ABM state and the normal to interface.

\begin{figure}[htbp]
\centering
	\includegraphics[width=1.0\linewidth]{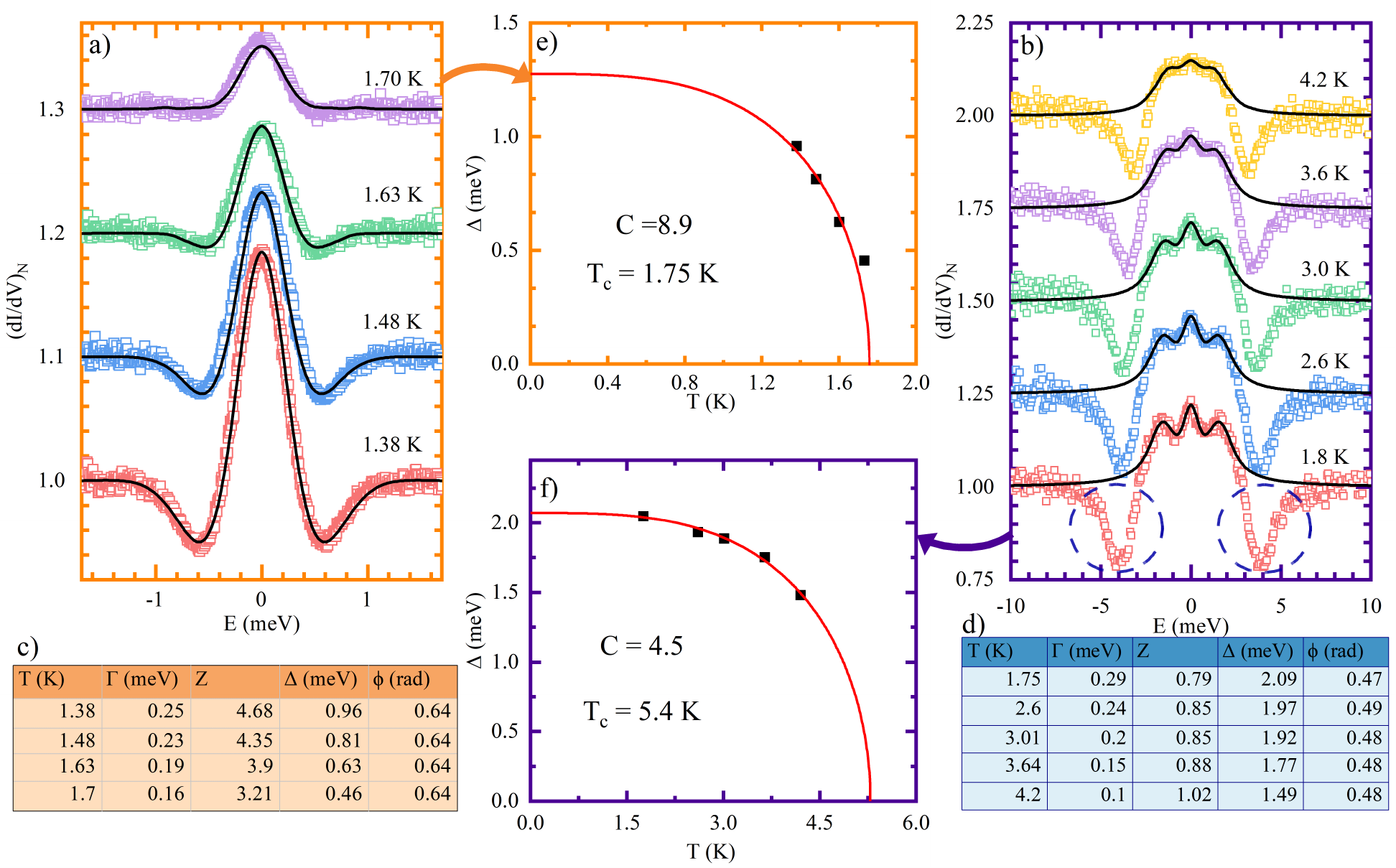}
\caption{\label{fig:PCS_ABM}
	Point contact spectroscopy data. a,b) Normalized PCS results obtained for 2 samples with the fitted  ABM model. 
	Dips at $ |E |>2.5$ $meV$ in b) are due to critical current \cite{sheet_role_2004} and stay beyond the employed model. The three peaks at lower energies were fitted assuming spectrum normalization at 10 $meV$, and reflect anisotropic order parameter
		 e), f) Superconducting order parameter $\Delta$ for both samples, obtained with  Eq. \ref{zgirski}. The evaluated critical temperature coincides with the temperature, at which the spectra appear. The value of the constant $C=\frac{\Delta_{T=0}}{k_bT_c}\gg1.76$, points to non BCS behavior.
		 c),d) summary of fitting parameters: table c) corresponds to panels a) and e),
		 table d) corresponds to panels b) and f).
		 }
\end{figure}

Experimental results of point-contact spectroscopy as a function of temperature for two different samples are presented in Figure \ref{fig:PCS_ABM}a) and b). The data are fitted with the theoretical model mentioned above. The model reproduces the data very well. The fit parameters are listed in Figure \ref{fig:PCS_ABM} c) and d). For given sets of point contact spectra for multiple temperatures, the values $\phi$ parameters are nearly constant for a given realization of contact.  The most external dips, marked on Figure \ref{fig:PCS_ABM}b) and attributed to the critical current \cite{sheet_role_2004}, are not taken into account due to limitations of the model. The spectra are smoother and have more easily distinguished features than the ones presented in Fig. \ref{fig:intropcs}. We link this fact to the presence of a varying size of the superconducting island-like regions identified in Ref. \cite{yuzephovich_interfacial_2008}.

Values of $\Delta$ obtained from the ABM fit in Fig. \ref{fig:PCS_ABM}a) and b) are presented in panels e) and 
f), respectively. Solid lines are fit of a simplified BCS formula \cite{Zgirski2008ExperimentalSO}:
\begin{equation}\label{zgirski}
    \Delta(T)=\Delta_{T=0}\left(1-\left(\frac{T}{T_c}\right)^{3.2}\right)^{0.5},
\end{equation}
where $\Delta_{T=0}$ is the SC order parameter $\Delta$ at absolute zero.  Both samples exhibit a much larger constant $C=\frac{\Delta_{T=0}}{k_bT_c}$ than the one predicted by BCS theory.  The departure from $C=1.76$, obtained for weak coupling limit, is usually associated with non-BCS superconductivity \cite{unconentional}. 

\subsubsection*{Four-probe transport characterization}

\begin{figure}[htbp]
\centering
\includegraphics[width=0.7\linewidth]{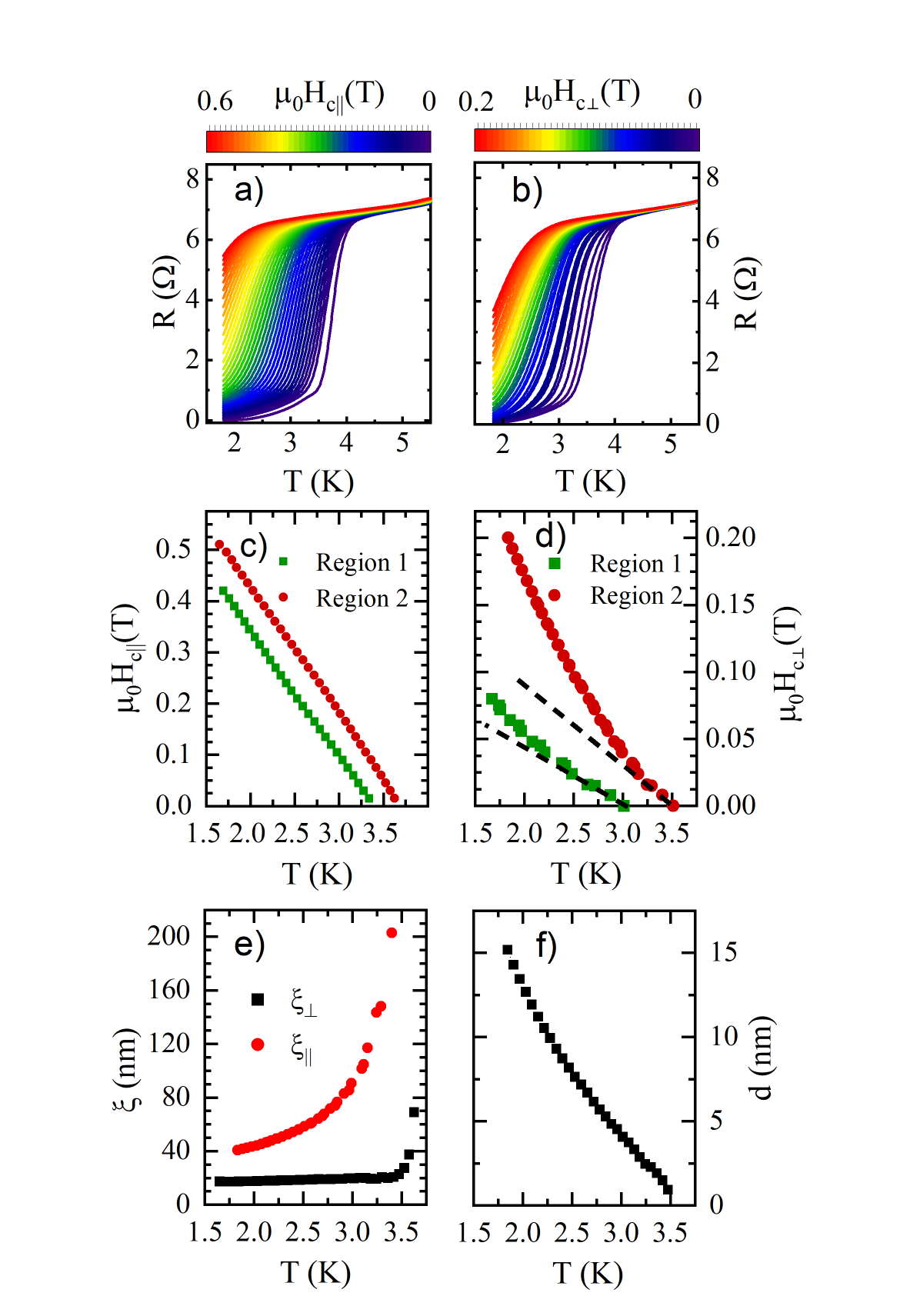}
\caption{\label{fig:xi} 
	Analysis of critical magnetic field.
	a) and b) The raw data of the four--probe resistance versus temperature  used for determination critical magnetic field,	panel a) in-plane  $\mu_0 H_{c2||}$ configuration, panel b) out-of-plane $\mu_0 H_{c2\perp}$ configuration
	 c)  In-plane critical magnetic fields determined in two different sample regions (denoted 1 and 2)
	 d)	 Out-of-plane critical magnetic fields determined in the same regions as in panel c)   A dashed straight line follows the linear dependence of the critical out-of-plane field for temperatures in  the vicinity of  $T_c$. They were drawn as a guide for the eye to show the kinks – see text.
			 e) perpendicular $\xi_{\perp}$ and parallel $\xi_{||}$ coherence lengths as a function of temperature obtained from Region 2 in panels c) and d).
			 f) The thickness of the superconducting layer $d$ determined from the critical magnetic fields.
			  }
\end{figure}

To shed more light on the system studied  we have conducted complementary transport experiments as a function of temperature and magnetic field (raw data are shown in Figure \ref{fig:xi}a,b)). Figures \ref{fig:xi}c) and d) show the evaluated critical magnetic field $\mu_0 H_{c2}$, in both configurations: in-plane $\mu_0 H_{c||}$  and out-of-plane $\mu_0 H_{c\perp}$ respectively. The measurements were done for a set of magnetic fields while sweeping the temperature in the range $1.5$ - $9$ K. The critical temperature is attributed to the point where the sample resistance reaches half of the normal state value. 

The temperature dependence of the out-of-plane  critical magnetic field presented in Figure \ref{fig:xi}d)  shows a distinct kink behavior. Similar  anisotropy of the critical magnetic field with the kink at least in one direction was observed in $UTe_2$ \cite{Ran_UTe2_2019}, $CeRh_2As_2$ \cite{Khim_CeRhAs2021} and $\alpha-BiPd ~~ 
\beta-Bi_2Pd$~ \cite{Chiang_unequivocal_2023}, where triplet pairing has been identified. 

Singlet superconductors exhibit upper critical field governed by orbital and paramagnetic pair breaking effects. A rough estimation of the
 orbital limit   may be made  within the Werthamer-Helfand-Hohenberg theory $H_{orb} = 0.7~~T_c (-(dH_{c2}/dT))_{T_c} $   \cite{WHH_1966}. It gives  $H_{orb\parallel}=0.64~T$   and  $0.77~T$ for Figure \ref{fig:xi}c)  and $H_{orb\perp}=0.10$ and $0.13~T$ for region 1 and 2 in Figure \ref{fig:xi}d) respectively.

 We also calculate the zero temperature Pauli limit by $H_{para} = 1.86~~T_c$ \cite{Clogstone_1962}, what gives $H_{para\parallel}= 6.2~T$ and $6.7~T$ for Fig. \ref{fig:xi}c)  ~~and~~
  $H_{para}=5.6$ and $6.0~T$ for region 1 and 2 in Figure \ref{fig:xi}d).
  In all presented cases $H_{para}> H_{orb}$, but due to a characteristic upward deflection (kink) our $H_{c \perp}$ data always exceed estimated orbital limit of pair breaking pointing to unconventional superconductivity, still below the Pauli limit. This behavior has been identified as spin-triplet SC in Ref.\cite{Ran_UTe2_2019,Khim_CeRhAs2021,Chiang_unequivocal_2023}.
  
 Having both critical fields determined, we calculate the Cooper pair coherence length $\xi_{\perp(||)}$ within the anisotropic Landau-Ginzburg equations \cite{wawro_electron_1993,nakajima_superconducting_1989}:

\begin{equation}\label{xi}
    \mu_0 H_{c\perp}=\frac{\Phi_0}{2\pi \xi_{||}^2} \quad ;
    \quad \mu_0 H_{c\parallel}=\frac{\Phi_0}{2\pi \xi_{||} \xi_{\perp}}
\end{equation}

The calculated values of coherence lengths do not differ much from dataset to dataset (regions 1-2), and the qualitative temperature trend is conserved. The results are presented in Figure \ref{fig:xi}e). One can see that $\xi_{\perp}$ never exceeds the thickness of the PbTe or SnTe layer ($100$ nm), which suggests Cooper pairs are localized at the single interface. On the other hand $\xi_{||}$  can reach hundreds of nanometers near the critical temperature. At low temperatures, the coherence length approaches a value of a few tens of nm, equal to around $10-20$ nm in the case of $\xi_{\perp}$ and $30-40$ nm in the case of $\xi_{||}$. Other quantities describing the SC state were evaluated:  the effective thickness of the superconducting phase $d$ and a dimensionless factor describing the effective strength of the two-dimensionality of the superconducting layer $r$ \cite{wawro_electron_1993,nakajima_superconducting_1989,fogel_interfacial_2002}: 

\begin{equation}\label{d_r}
    d^2=\frac{6\Phi_0H_{c\perp}}{\pi \mu_0 H_{c\parallel}^2}  \quad ; \quad r=\frac{4}{\pi}\left(\frac{2\xi_\perp}{\lambda}\right)^2,
\end{equation} where $\lambda$ is the distance between SC interfaces ($\lambda\simeq100$ nm). The calculated value of $d$ as a function of temperature is presented in Figure \ref{fig:xi}f). This parameter increases with decreasing temperature, reaching $15~nm$ at $T=1.5~K$. This finding agrees with the predictions of Tang and Fu \cite{tang_strain-induced_2014}: as stress from interface dislocation relaxes far from the interface, the $T_c$ can be spatially dependent along the growth direction. This fact effectively translates to an increase of superconducting phase volume with decreasing temperature.
The interpretation of the $r$ parameter introduced above is as follows: when $r\ll1$, the SC phase is 2D; when $r\gg1$, the SC phase is 3D. The evaluated $r$ is consistently smaller than $0.25$, explicitly confirming the 2D nature of SC condensate \cite{nakajima_superconducting_1989}. 

\subsubsection*{Magnetic studies}

\begin{figure}[htbp]
\centering
\includegraphics[width=0.7\linewidth]{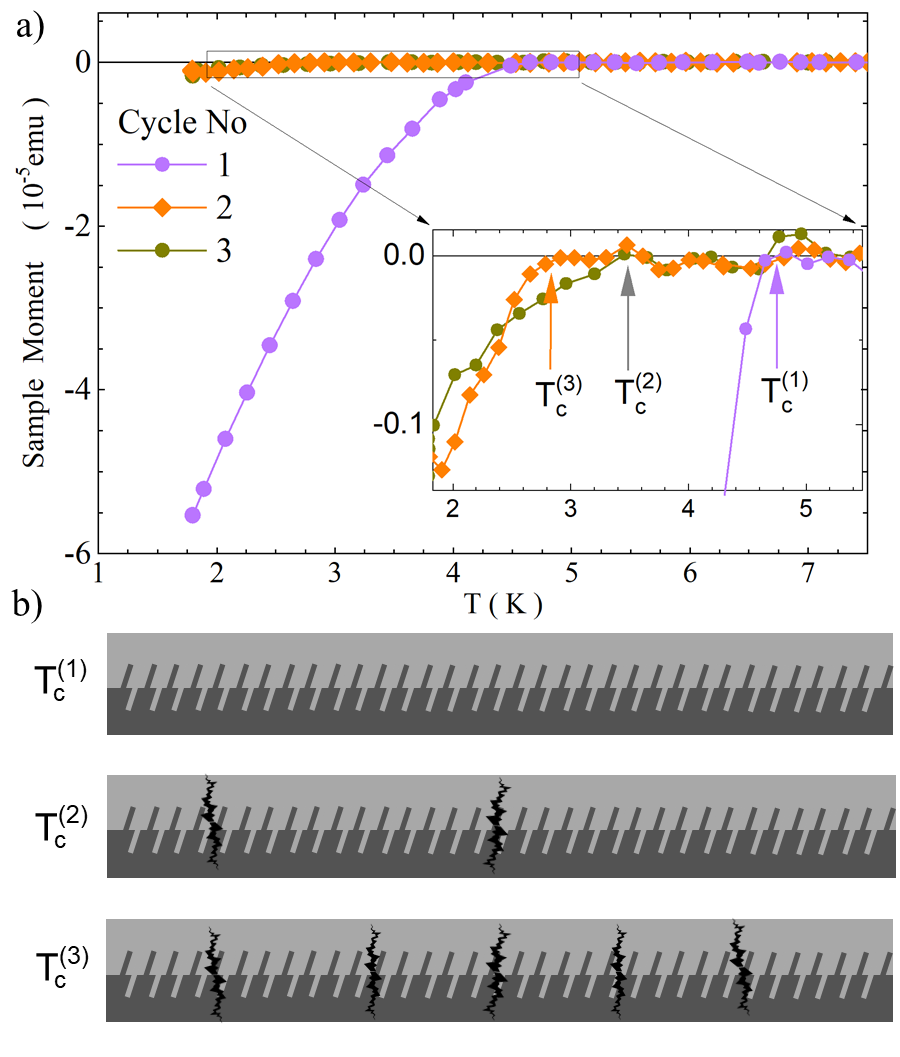}
\caption{\label{fig:squid}
	Degradation of superconductivity during thermal cycles. a) Temperature dependence of magnetization measured in 10 Oe for 3 subsequent cooling processes, showing the Meissner effect fading out. b) Scheme of possible sample degradation of the dislocation grid on the interface, leading to a loss of periodicity and deterioration of 
superconducting properties.}
\end{figure}

Magnetization studies were performed using SQUID magnetometry  at temperatures from $300$ K to $1.8$ K. Several consecutive cooling cycles (Figure \ref{fig:squid}a) ) have evidenced clear changes in magnetic character of the sample: (i) the investigated structure exhibits the Meissner effect, and (ii) that the onset of the diamagnetic signal shifts to lower temperatures, indicating the change in $T_c$. This change is related to degradation of the periodic dislocation grid in subsequent cooling processes \cite{fogel_interfacial_2002}, schematically shown in Figure \ref{fig:squid}b). We may thus exclude that the superconductivity of the samples originates from grainy precipitates of superconducting elements (Pb or Sn) \cite{mazur_2019_experimental,darchuk_1998_phase}, because its magnetic state would stay insensitive to the thermal cycling of the sample \cite{Eichele1981-al,akselrod_superconductivity_1974}.

\subsection*{Conclusions}

This paper presents set of observations which indicate spin-triplet pairing in the MBE-grown PbTe/SnTe semiconductor heterostructures. Point-contact spectroscopy with spin polarized and unpolarized electrons exhibits qualitatively similar spectra with a zero bias peak, in a stark contrast to the results obtained in a control experiment with singlet superconductor (Nb), where spin-polarized spectrum was remarkably suppressed at zero bias. The point contact spectra were successfully fitted with the Anderson-Brinkman-Morel model of spin-triplet superconductivity. Additionally, the four probe resistivity measurements allowed us to extract the in-plane and out-of-plane critical magnetic fields and follow their temperature evolution. Our results revealed a characteristic kink behavior of perpendicular critical magnetic field  on temperature, which has recently been reported to be unequivocal identification of spin-triplet superconducting phase. Our data corroborate with the predictions of Tang and Fu \cite{tang_strain-induced_2014} offering a novel way of designing flatband systems by means of material and strain engineering.


\newpage
\clearpage
\section*{Author Information}
\subsection*{Corresponding Authors}
{\bf P. Sidorczak}\ \\
{\small University of Warsaw, Faculty of Physics, ul. Pasteura 5, 02-093, Warsaw, Poland.\ \\
International Research Centre MagTop, Institute of Physics, Polish Academy of Sciences, \ \\ al. Lotnikow 32/46, 02668,  Warsaw, Poland.\ \\
Center for Quantum Devices, Niels Bohr Institute, University of Copenhagen,  Universitetsparken 5, 2100 Copenhagen, Denmark.\ \\
E-mail: pawel.sidorczak@nbi.ku.dk}
\ \\
{\bf M. Gryglas-Borysiewicz}\ \\
{\small University of Warsaw, Faculty of Physics, ul. Pasteura 5, 02-093, Warsaw, Poland.\ \\
E-mail: Marta.Gryglas@fuw.edu.pl}	
\ \\
{\bf K. Dybko}\ \\
{\small Institute of Physics, Polish Academy of Sciences,   al. Lotnikow 32/46, 02668,  Warsaw, Poland. \ \\
International Research Centre MagTop, Institute of Physics, Polish Academy of Sciences,\ \\ al. Lotnikow 32/46, 02668,  Warsaw, Poland.}

\subsection*{Notes}
The authors declare no competing financial interest

\begin{acknowledgement}
 This study has been supported by the National Science Centre (Poland) through OPUS (UMO - 2017/27/B/ST3/02470) project and by the Foundation of Polish Science through the IRA Programme co-financed by EU within SG OP.
\end{acknowledgement}

\begin{suppinfo}
The file SupportingInfo-Sidorczak25.pdf is available free of charge.
\ \\
Materials and methods, Experimental methods, Structural characterization,\   \\
Methods of Sample preparation for transport measurements,\ \\
Extended data of PCS charaterization, \ \\
Control experiment with s-type SC Niobium, \ \\
Four-probe resistance dependence on temperature, \ \\
Microscopic studies  on elemental content of heterostructures, \ \\
Broken Pauli limit and record Tc for PbTe/SnTe heterostructure,  \ \\
Samples, \ \\
References \ \\
\end{suppinfo}

\bibliography{Sidorachem}

\end{document}